\documentclass[11pt]{article}

\usepackage{amsmath,amssymb}
\usepackage{pifont,fancybox}
\usepackage{graphicx}
\usepackage{psfrag}
\usepackage{epsfig,rotate}
\usepackage{umlaut}
\usepackage{bpj}
\usepackage{fancyheadings}

\def\thebibliography#1{\section*{References}
  \list
  {\relax}{\setlength{\labelsep}{0em}
    \setlength{\itemsep}{-\parsep}
    \setlength{\itemindent}{-\bibhang}
    \setlength{\leftmargin}{\bibhang}\small}
  \def\newblock{\hskip .11em plus .33em minus .07em}
  \sloppy\clubpenalty4000\widowpenalty4000
  \sfcode`\.=1000\relax}

\begin{document}

\baselineskip = 0.9\baselineskip
\addtolength{\textfloatsep}{-\baselineskip}

\def\gvect#1{\mbox{\boldmath $#1$}}
\newcommand{\hamil}{{\cal H}}
\newcommand{\pp}[2]{\frac{\partial #1}{\partial #2}}
\newcommand{\ppp}[3]{\left.\frac{\partial #1}{\partial #2}\right|_{#3}}
\newcommand{\avg}[1]{\left\langle #1 \right\rangle}
\newcommand{\dOne}[1]{\!\!d#1\,}
\newcommand{\dTwo}[1]{\!\!d^2#1\,}
\newcommand{\dThree}[1]{\!\!d^3#1\,}
\newcommand{\dd}[2][{}]{\!\!d^{#1}#2\,}
\newcommand{\dPath}[1]{\!\!{\cal D}\!\left[#1\right]\,}
\newcommand{\lang}{{\cal L}}
\newcommand{\vabs}[1]{{\left|#1\right|}}
\newcommand{\vect}[1]{{\mathbf #1}}
\newcommand{\etal}{et~al.\ }
\newcommand{\eg}{e.\,g.}
\newcommand{\etc}{etc.\ }
\newcommand{\order}[1]{{\cal O}(#1)}
\newcommand{\mum}{\ensuremath{\mu\text{m}}}
\newcommand{\rem}[1]{{\bf Remark: #1}}
\renewcommand{\thefootnote}{\fnsymbol{footnote}}

\newcommand{\shorttitle}{Statistical mechanics of semiflexible polymers}
  \lhead[\sl\thepage]{\sl\shorttitle}
\rhead[\sl\shorttitle]{\sl\thepage} \cfoot{}

\setcounter{page}{1}
\pagestyle{plain}

\begin{center}{\Large\bf  LECTURE 9 

 \vspace{1cm}

    Statistical mechanics of semiflexible polymers: \\ theory and
    experiment\footnote{\em To be published in ``Dynamical networks in physics
      and biology'' (Proceedings of a Les-Houches workshop), edited by G.
      Forgacs and D. Beysens, EDP Sciences Springer Verlag
      (1998).}}\end{center}

\begin{center}
  {Erwin Frey, Klaus Kroy, Jan Wilhelm and Erich Sackmann} 
\end{center}
\begin{center}
  {\sl Institut f\"ur Theoretische Physik und Institut f\"ur Biophysik,
    Physik-Department der Technischen Universit\"at M\"unchen,
    James-Franck-Stra\ss e, D-85747 Garching, Germany}
\end{center}

\centerline{ ------------------------------ } 

\vspace{-0.2cm}

\centerline{ -------------------- }

\vspace{1cm}

\section{Introduction and overview}

Living cells are soft bodies of a characteristic form, but endowed with a
capacity for a steady turnover of their structures. Both of these material
properties, i.e.\ recovery of the shape after an external stress has been
imposed and dynamic structural reorganization, are essential for many cellular
phenomena. Examples are active intracellular transport, cell growth and
division, and directed movement of cells. The structural element responsible
for the extraordinary mechanical and dynamical properties of eukaryotic cells
is a three-dimensional assembly of protein fibers, the {\em cytoskeleton}. A
major contribution to its mechanical properties is due to actin filaments and
proteins that crosslink them. Numerous experiments {\it in vivo}
\cite{eichinger-etal:96} and {\it in vitro} \cite{janmey:91,janmey:95} have
shown that the mechanical properties of cells are largely determined by the
cytoskeletal network.

\begin{figure}[tb]
\begin{center}
 \psfrag{ext}[c]{\small $k_BT\delta R/fL^2$}
\psfrag{lp}{\small $\ell_p/\! L$}
\centerline{\epsfxsize=1.2truein \epsffile{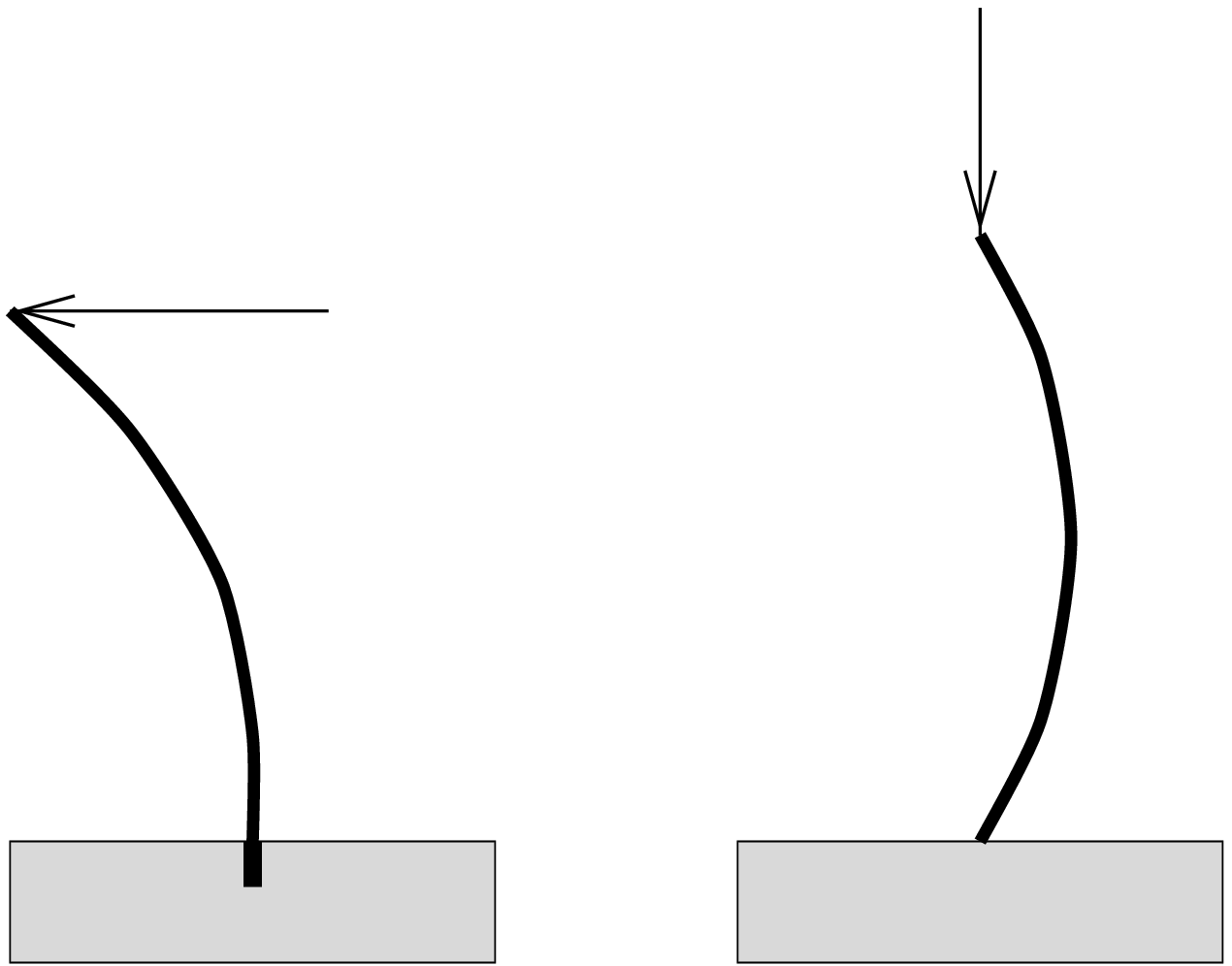}
\hspace{1cm} \epsfxsize=1truein \rotate[r]{\epsffile{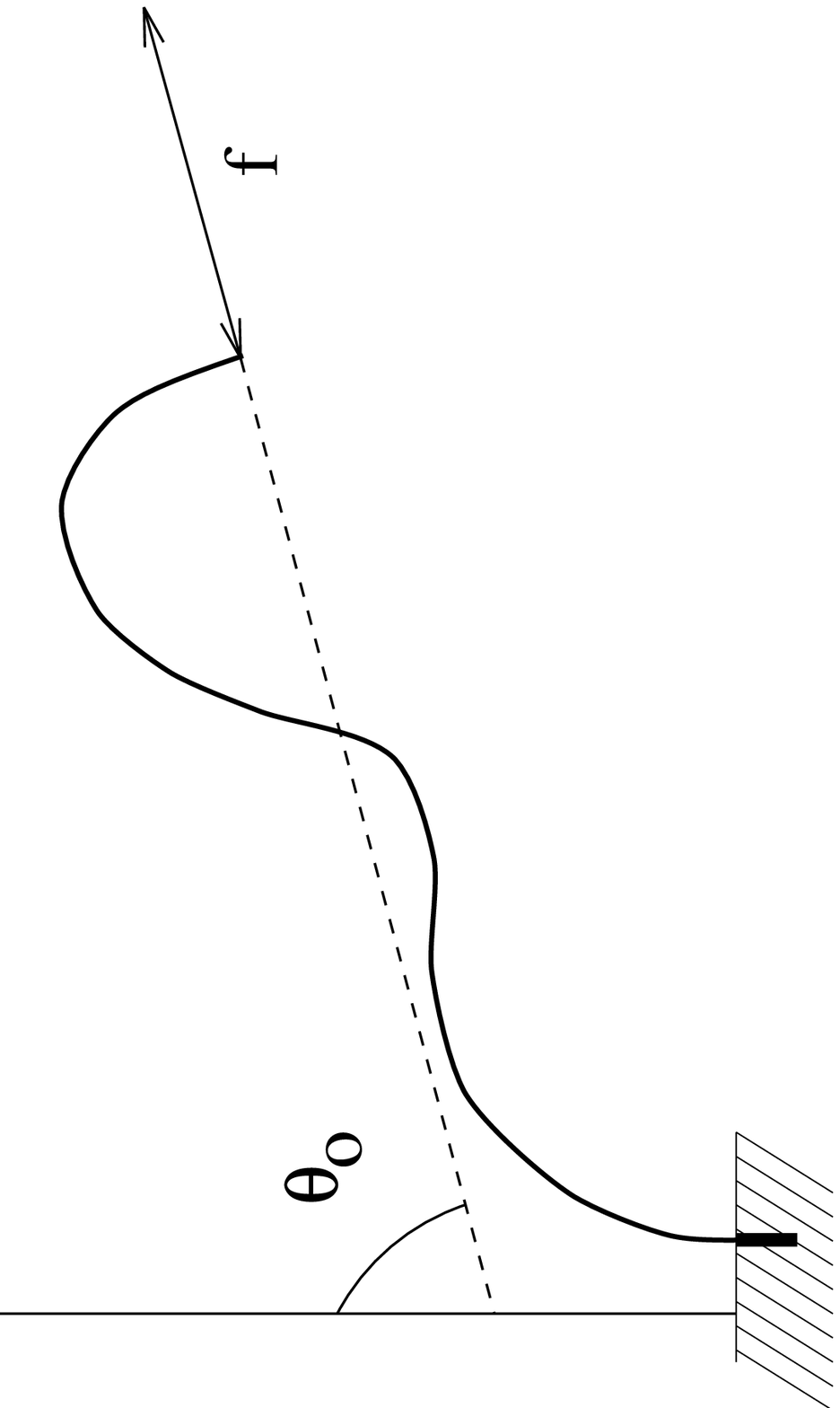}}}
\centerline{\epsfxsize=2.4truein \epsffile{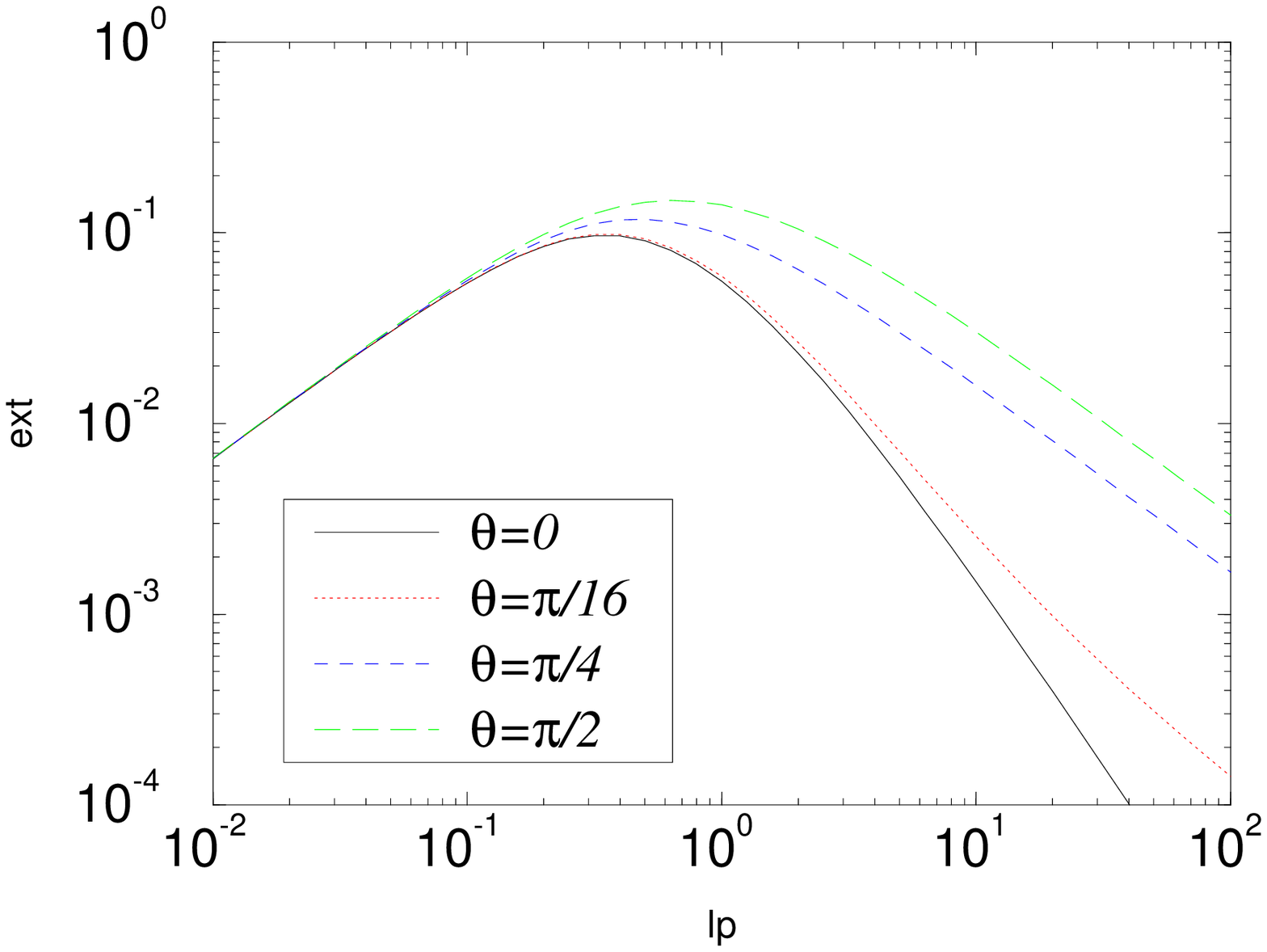}}
\caption{\label{fig:force_extension}
  \small \emph{Top:} The elastic response of a stiff rod is extremely
  anisotropic due to the Euler instability. \emph{Bottom:} Response of a
  filament clamped at one end with a fixed initial orientation to a small
  external force at the other end. The normalized extension (inverse force
  coefficient) is plotted versus the persistence length in units of the total
  length of the polymer. The curves are parameterized by the relative angle
  $\theta_0$ of the force ${\bf f}$ to the initial orientation.}
\end{center}
\end{figure}

The cytoskeletal polymers (actin filaments, microtubules and intermediate
filaments) which build up this network are at the relevant length-scales (a few
microns at most) all {\it semiflexible polymers}. In contrast to flexible
polymers the {\em persistence length} of these polymers is of the same order of
magnitude as their total contour length. Hence, the statistical mechanics of
such macromolecules can not be understood from the conformational entropy alone
but depends crucially on the bending stiffness of the filament.  The nontrivial
elastic response and distribution function of the end-to-end distance of a
single semiflexible polymer illustrate instructively this interplay of
energetic and entropic contributions.  Fig.~\ref{fig:force_extension} shows the
extension (in units of $fL^2/ k_B T$) of a semiflexible polymer of fixed length
$L$ grafted at one end when a weak force $f$ is applied at the other end for
different persistence lengths $\ell_p$ \cite{kroy-frey:96}. This is one of the
few {\em exact} results for the wormlike chain model.  Note that the largest
response is obtained for a polymer with a persistence length of the order of
its contour length. More flexible molecules contract due to the larger entropy
of crumpled conformations, stiffer molecules straighten out to keep their
bending energy low.  In the flexible case the response is isotropic and
proportional to $1/k_BT$, i.e., the Hookian force coefficient is proportional
to the temperature and we recover ordinary rubber elasticity. On the other
hand, when the persistence length is longer than the contour length, the
response becomes increasingly {\em anisotropic}.  Transverse forces give rise
to mechanical bending of the filaments and the transverse spring coefficient in
the stiff limit is proportional to the bending modulus $\kappa$.  The effective
longitudinal spring coefficient turns out to be proportional to $\kappa^2/T$,
indicating the breakdown of linear response at low temperatures $T\to0$ (or
$\ell_p\to\infty$).  This is a consequence of the well known Euler buckling
instability illustrated in Fig.~\ref{fig:force_extension} (top). The anisotropy
of the elastic response considerably complicates the construction of
network models.

\pagestyle{fancy}

Energetic effects also have a strong influence on the probability distribution
function $G(R; L)$ for the end-to-end distance $R$ of a semiflexible polymer of
length $L$. When the polymer is not much longer than its persistence length,
$G(R; L)$ deviates strongly from the Gaussian shape found for flexible polymers
(see Fig.~\ref{fig:verteilungsfunktion}) with the main weight being shifted
towards full extension with increasing stiffness \cite{wilhelm-frey:96}. This
is also reflected by the nonlinear force-extension relation, whose overall
shape is determined by the Euler instability.  Measurements of the radial
distribution function as well as of the response of single polymers to applied
forces can be used to measure the bending stiffness of biopolymers.

A useful experimental technique for investigating the short time dynamics of
semiflexible polymers is dynamic light scattering
\cite{farge-maggs:93,harnau-winkler-reineker:96,kroy-frey:97}.  We have
calculated the dynamic structure factor for solutions of semiflexible polymers
in the weakly bending rod limit. The result allows characteristic parameters of
the polymers, such as their persistence length $\ell_p$, their lateral diameter
$a$ and also the mesh size $\xi_m$ of the network to be determined from light
scattering measurements.  The most important result of this work is that the
decay of the dynamic structure factor shows two regimes, an initial simple
exponential decay determined by the hydrodynamics of the solvent and a
stretched exponential decay due to structural relaxation driven by the bending
energy.  For actin and intermediate filaments, where both regimes can be
resolved, the method is well suited to investigate also more complex questions
concerning for example the interactions with actin binding proteins which among
others can induce cross-linking or bundling.

In order to describe the material properties of the cytoskeleton, one has to
understand how semiflexible polymers built up statistical networks and how the
macroscopic stresses and strains are mediated to the single filaments in such a
``rigid polymer network''. This is a quite rapidly growing field in polymer
physics with many questions still open \cite{aharoni-edwards:94}.

If the polymers are not crosslinked, the response will depend on how fast one
pulls.  Roughly speaking, the solution will either show elastic behavior and
obey Hook's law with a linear relation $\sigma = G \gamma$ between stress
$\sigma$ and strain $\gamma$ or rather show viscous behavior and obey Newton's
law $\sigma = \eta {\dot \gamma}$, where the stress is proportional to the
strain rate ${\dot \gamma}$.  The task of a theoretical description is to find
out how the material parameters like the shear modulus $G^0$ and the viscosity
$\eta$ depend on the elastic and dynamic properties of the polymers and the
architecture of the network. The mechanical and dynamical properties of the
various crosslinking proteins are also expected to influence the
viscoelasticity of the network \cite{wachsstock-schwarz-pollard:94}.

Whereas it is known that on short time scales the effect of mutual steric
hindrance (``entanglement'') in conventional polymeric materials is very
similar to the effect of permanent chemical crosslinks, this is an open problem
for semiflexible polymer networks. Since forces between neighboring polymers
can only be transmitted transverse to the polymer axis and there is no
restoring force for sliding of one filament past another, a steric contact is
acting completely different from a chemical crosslink on a microscopic scale.

But even the most simple of these problems, the static elasticity of a network
of permanently crosslinked semiflexible polymers, is quite complex. Because of
the strongly anisotropic behavior of the single elements, the predicted
macroscopic properties of the network vary greatly with the explicit or
implicit assumptions made about network geometry and stress propagation by
recent theoretical treatments of both entangled solutions and crosslinked
networks \cite{mackintosh-kaes-janmey:95,isambert-maggs:96,%
  satcher-dewey:96,kroy-frey:96}.  We hope to represent a key aspect of the
geometrical structure of both cellular and artificial stiff polymer networks by
looking at \emph{disordered} networks. Specifically, we use a crosslinked
network of sticks randomly placed in a plane as a toy model for studying the
origin of macroscopic elasticity in a stiff polymer network. Although
quantitative predictions about the behavior of existing (three-dimensional)
networks of semiflexible polymers are not attempted at this stage, this model
is expected to reflect the salient features of the full problem and to promote
its understanding by allowing the detailed discussion of questions like ``What
modes of deformation contribute most to the network elasticity?'', ``How many
filaments do actually carry stress, how many remain mostly unstressed?'',
``What kind of effective description of the complicated microscopic network
geometry should be used?''.  This approach connects the theory of cytoskeletal
elasticity to the very active fields of transport in random media and elastic
percolation. In section~\ref{stochastic_network_models} we will describe the
model in more detail and discuss the question of the dominant deformation mode.

Previous fluorescence microscopic observations \cite{kaes-strey-sackmann:94}
suggest that the tube picture is a useful concept for understanding the
viscoelastic properties of \emph{entangled} semiflexible polymer networks.
Starting from this model, where the matrix surrounding a single test polymer is
represented by a tube-like cage, we have developed a phenomenological
description that seems to be able to account for some of the observed elastic
and dynamic properties. The predictions for the entanglement transition, the
concentration dependence of the plateau modulus, and the terminal relaxation
time are in good quantitative agreement with experimental data.


\section{Single-chain properties}

\subsection{The wormlike chain model}
The theoretical understanding of the mechanical properties of a {\em single}
semiflexible macromolecule in isolation is already a nontrivial statistical
mechanics problem with quite a number of recent developments 50 years after it
was first formulated \cite{kratky-porod:49}. The model usually adopted for a
theoretical description of semiflexible chains like actin filaments is the {\em
  wormlike chain model}. The filament is represented by an inextensible space
curve ${\bf r}(s)$ of total length $L$ parameterized in terms of the arc length
$s$.  The statistical properties of the wormlike chain are determined by a free
energy functional which measures the total elastic energy of a particular
conformation.
\begin{eqnarray}
\label{eq:hamiltonian}
{\cal H}  = \int_0^L ds \, \, \frac{\kappa}{2} 
            \left( \frac{\partial {\bf t}}{\partial s}\right)^2; \quad
             | {\bf t} | = 1 \, ,
\end{eqnarray} 
where ${\bf t} (s) = \partial {\bf r} (s) / \partial s$ is the tangent vector.
The energy functional is quadratic in the local curvature with $\kappa$ being
the bending stiffness of the chain. The inextensibility of the chain is
expressed by the local constraint, $|{\bf t} (s)|=1$, which leads to
non-Gaussian path integrals. Since there would be high energetic costs for a
chain to fold back onto itself one may safely neglect self-avoidance effects
for sufficiently stiff chains.

Despite the mathematical difficulty of the model some quantities can be
calculated exactly.  Among these is the tangent-tangent correlation function
which decays exponentially, $\avg{ {\bf t} (s) \cdot {\bf t} (s')} = \exp
\left[ - (s-s') / \ell_p \right]$, with the persistence length
$\ell_p=\kappa/k_BT$ (in three dimensional space). Another example is the
mean-square end-to-end distance
\begin{eqnarray}
{\cal R}^2 &:=&    \langle R^2 \rangle = 2\ell_p^2(e^{-L/\ell_p}-1+L/\ell_p)
\nonumber \\ &=&
  \begin{cases}
    L^2 &\text{ for } L/\ell_p \rightarrow 0  \text{ (rigid rod)} \\
    2 \ell_p L &\text{ for } L/\ell_p \rightarrow \infty \text{ (random
      coil)},
\end{cases}
\end{eqnarray}
which reduces to the appropriate limits of a rigid rod and a random
coil (with Kuhn length $2\ell_p$) as the ratio of $L$ to $\ell_p$
tends to zero or infinity, respectively.  The calculation of higher
moments quickly gets very troublesome
\cite{hermans-ullman:52}.

\subsection{Linear force-extension relation}\label{sec:linear_response}
Another useful property of the model, which can be computed exactly is the
linear force extension relation \cite{kroy-frey:96}.  Adding a term $-\vect{f}
\cdot \vect{R}$ to the Hamiltonian in Eq.\ \ref{eq:hamiltonian}, the extension
of a wormlike chain with a weak force applied between its ends is computed as
\begin{equation}
  \label{eq:extension}
 \delta R= \left( \langle {\bf R} \rangle_f - \langle {\bf R} \rangle_0
  \right) {\bf f}/f = f ({\cal R}^2-\avg{|\vect R|}^2)/k_bT  = f/k  
\end{equation}
with $k$ the force coefficient. The problem is the calculation of the moment of
the the end-to-end distance $\avg{|\vect R|}$. This has not been achieved so
far, but an expansion in the stiff limit is possible. For example, we can write
$\avg{|\vect R|}={\cal R}\langle\!\sqrt{1+( R^2-{\cal R}^2)/\!{\cal
    R}^2}\rangle$ and expand the square root to obtain to leading order for the
force coefficient
\[
4k_BT\frac{{\cal R}^2}{\avg{ R^4}-{\cal R}^4}
\xrightarrow{L/\ell_p\ll1} \frac{90\kappa^2}{k_BT L^4} .
\] 
This is exact in the stiff limit, $L/\ell_p\ll1$, and qualitatively correct
over the whole range of stiffness.  However, for the special boundary condition
of a grafted chain we can even get an {\em exact result for arbitrary
  stiffness}.  Consider a chain with one end clamped at a fixed orientation and
let us apply a force at the other end. Then the linear response of the chain
may be characterized in terms of an effective Hookian spring constant which
depends on the orientation $\theta_0$ of the force with respect to the tangent
vector at the clamped end. In the appropriate generalization of the second part
of Eq.\ \ref{eq:extension} the force coefficient $k$ is replaced by an angle
dependent function, $k^{-1}\to L^2{\cal F}_{\theta_0}(\ell_p/L) /k_BT$.  Noting
that the conformational statistics of the wormlike chain is equivalent to the
diffusion on the unit sphere \cite{saito-takahashi-yunoki:67} the function
$\cal F$ can be calculated \cite{kroy-frey:96} (see
Fig.~\ref{fig:force_extension}).  In the flexible limit, where the chain
becomes an isotropic random coil, all curves coincide and reproduce entropy
elasticity, ${\cal F}_{\theta_0}(x)\sim x$.  But for stiff chains the
force-extension relation depends strongly on the value of the angle $\theta_0$
between the force and the grafted end.  Transverse forces give rise to ordinary
mechanical bending characterized by the bending modulus $\kappa$ (${\cal
  F}_{\pi/2}\sim 1/x$), whereas longitudinal deformations are resisted by a
larger force coefficient $\kappa^2/T$ (${\cal F}_0\sim 1/x^2$).  The breakdown
of linear response in the limit $T\to0$ (or $\ell_p\to\infty$) is a consequence
of the well known Euler buckling instability, as we already mentioned in the
introduction. The linear response for longitudinal forces is due to the
presence of thermal undulations, which tilt parts of the polymer contour with
respect to the force direction.

\subsection{Nonlinear Response and radial distribution function}

In viscoelastic measurements on {\it in vitro} actin networks one observes
strain hardening \cite{janmey-etal:90}, i.e.\/ the system stiffens with
increasing strain. This may either result from collective nonlinear effects or
from the nonlinear response of the individual filaments. In the preceding
section we have seen that the force coefficient obtained in linear response
analysis for longitudinal deformation diverges in the limit of vanishing
thermal fluctuations indicating that the regime of validity for linear response
shrinks with increasing stiffness. Since the nonlinear response of a single
filament may be obtained from the radial distribution function by integration,
we discuss the latter first.

\begin{figure}[tb]
\begin{center}
  \vspace{0.2cm} \valign{\vfil#\vfil\cr
    \hbox{\includegraphics[width=.45\textwidth]{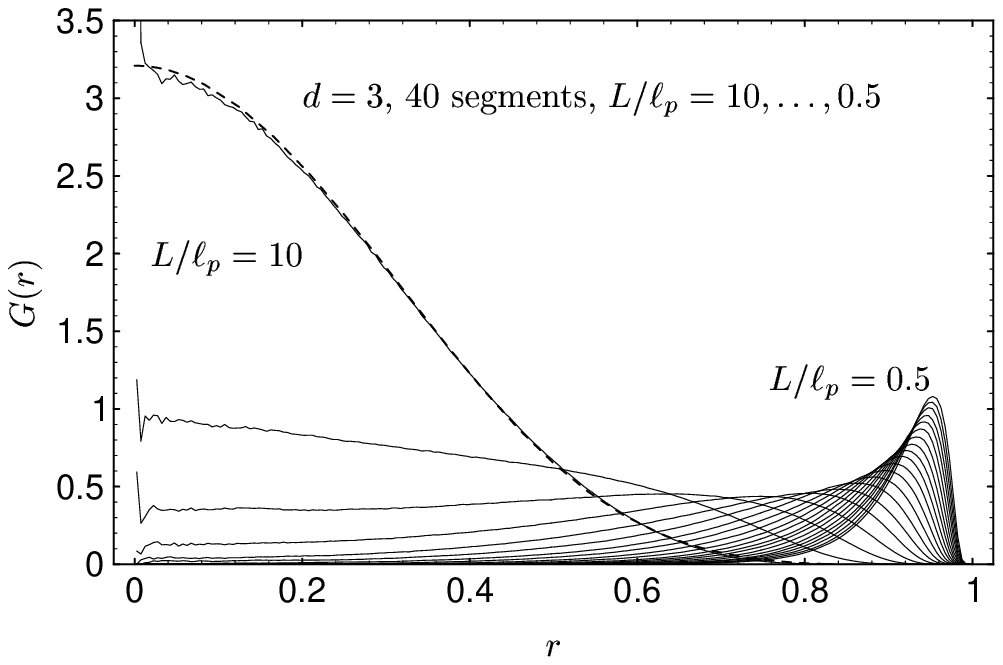}
      } \cr \hbox{\hspace{1cm}
      \includegraphics[width=.45\textwidth]{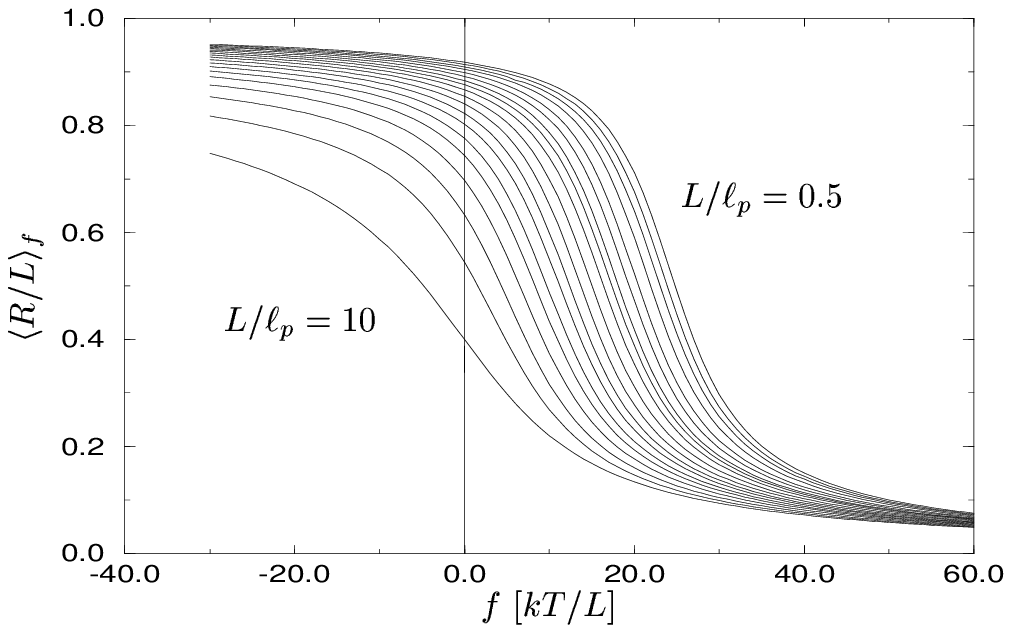}}
    \cr}
\end{center}

\vspace{-0.8cm}
\caption{\label{fig:verteilungsfunktion} \small {\it Left:} End-to-end 
  distribution function of a semiflexible polymer (numerical results). Note
  that with increasing stiffness of the polymer there is a pronounced crossover
  from a Gaussian to a completely non-Gaussian from with the weight of the
  distribution shifting towards full stretching. {\it Right:} The mean
  end-to-end distance $R$ as a function of a force applied between the ends
  ($\vect f = -f \vect R/|\vect R|$). The step at positive (i.e. compressive)
  forces can be viewed as a remnant of the Euler instability.}
\end{figure}

A central quantity for characterizing the conformations of polymers is the
radial distribution function $G({\bf R};L)$ of the end-to-end vector ${\bf R}$.
For a freely jointed phantom chain (flexible polymer) it is known exactly and
for many purposes well approximated by a simple Gaussian distribution. While
rather flexible polymers can be described by corrections to the Gaussian
behavior \cite{daniels:52}, the distribution function of polymers which are
shorter or comparable to their persistence length shows very different
behavior. It is in good approximation given by
\begin{eqnarray}
  \label{eqn:dist_func}
  &&G(\vect R; L) \approx \frac{\ell_p}{{\cal N} L^2} 
  f\Bigl(\frac{\ell_p}{L}(1-R/L)\Bigr), \nonumber \\
  &&\text{where } \; f(x) = 
  \begin{cases} \displaystyle 
    \tfrac{\pi}{2}  \exp[- \pi^2 x] 
    &\text{for } x > 0.2 \\
    \displaystyle
    \frac{1/x-2}{8\pi^{3/2} x^{3/2}} 
    \exp\left[-\frac{1}{4 x} \right] 
    &\text{for } x \le 0.2
  \end{cases}
\end{eqnarray}
and ${\cal N}$ is a normalization factor close to 1 \cite{wilhelm-frey:96}.
This result is valid for $L \lessapprox 2\ell_p$, $x \lessapprox 0.5$ and $d =
3$ where $d$ is the dimension of space. A similar expression exists for $d=2$.
As can be seen in Fig.~\ref{fig:verteilungsfunktion}, the maximum weight of the
distribution shifts towards full stretching as the stiffness of the chain is
increased to finally approach the $\delta$-distribution like shape required for
the rigid rod.

The radial distribution function is a quantity directly accessible to
experiment since fluorescence microscopy has made it possible to observe the
configurations of thermally fluctuating biopolymers
\cite{gittes-etal:93,kaes-etal:93,ott-magnasco-simon-libchaber:93}.  Comparing
the observed distribution functions with the theoretical prediction for $d=2$
is both a test of the validity of the wormlike chain model for actual
biopolymers as well as a sensitive method to determine the persistence length
which is the only fit parameter. It should be noted here
that the determination of persistence length e.g.  of actin is still an
actively discussed subject \cite{dupuis-guilford-warshaw:96,wiggins-etal:97}.

A very interesting possibility would be to attach two or more markers (e.g.,
small fluorescent beads) permanently to single strands of polymers and to
observe the distribution function of the marker separation.  This would
eliminate all the experimental difficulties associated with the determination
of the polymer contour. Note that in contrast to existing methods of analysis
it is not necessary to know the length of polymer between two markers; it can
be extracted from the observed distribution functions along with $\ell_p$ by
introducing $L$ as a second fit parameter.

The nonlinear response of the polymer to extending or compressing forces can be
obtained from the radial distribution function by integration. The result
(Fig.~\ref{fig:verteilungsfunktion}) is in agreement with and provides the
transition between the previously known limits of linear response and very
strong extending forces (e.g., \cite{marko-siggia:95}). For compressional
forces, a pronounced decrease of differential stiffness around the classical
critical force $f_c = \kappa \pi^2/L^2$ can be understood as a remnant of the
Euler instability. For filaments slightly shorter than their persistence length
the influence of this instability region extends up to and beyond the point of
zero force corresponding to the maximum in the linear response coefficient for
$\ell_p \approx L$ (see Fig.~\ref{fig:force_extension}). For large compressions
beyond the instability, the force-extension-relation calculated from the
distribution function is only in qualitative agreement with numerical results
because of the restricted validity of Eq.~\ref{eqn:dist_func} for $x \to 1$.

\subsection{Dynamic structure factor}

The short time dynamics of polymers is most effectively measured by scattering
techniques. The {\em dynamic structure factor} of a semiflexible polymers can
be derived analytically in the limit of strong length scale separation $a \ll
\lambda \ll \ell_p$, $L$ (`weakly bending rod' approximation) and $\lambda
\leq\xi_m$. Here we have introduced the symbols $a$, $\lambda$ and $\xi_m$ for
the lateral diameter of the polymer, the scattering wavelength and the mesh
size of the network, respectively. The length scale separation guarantees that
the decay of the structure factor is due to the {\em internal dynamics of
  single filaments}.  For $\lambda\approx\xi_m$ the structure factor decays in
two steps due to the fast internal and to the slower collective modes,
respectively. For the analysis we will restrict ourselves to rather dilute
solutions and refer the reader for a more complete treatment to the literature
\cite{kroy-frey:97}.

The most important result of the calculations is that the time decay of the
dynamic structure factor shows two regimes: the simple exponential initial
decay regime for $t\ll\tau= \tilde\zeta_\perp/\kappa k^4$
with a decay rate
\begin{equation}\label{eq:gamma_0}
  \gamma_k^{(0)}= \frac{k_BT}{6\pi^2 \eta}k^3\left(\frac56-\ln ka\right),
\end{equation}
 and the stretched exponential decay
\begin{equation}\label{eq:stretched_exp}
 g({\bf k},t) \propto 
              \exp
                  \left( -\frac{\Gamma(\frac14)}{3\pi}
                          (\gamma_kt)^{\frac34}
                  \right), \qquad 
 \gamma_k = k_BT k^{\frac83}/\ell_p^{1/3} {\tilde \zeta_{\perp}}
\end{equation}
for $t\gg\tau$. 
\begin{figure}[htb]
  \begin{center}
    \centerline{\epsfxsize=2.0truein \epsffile{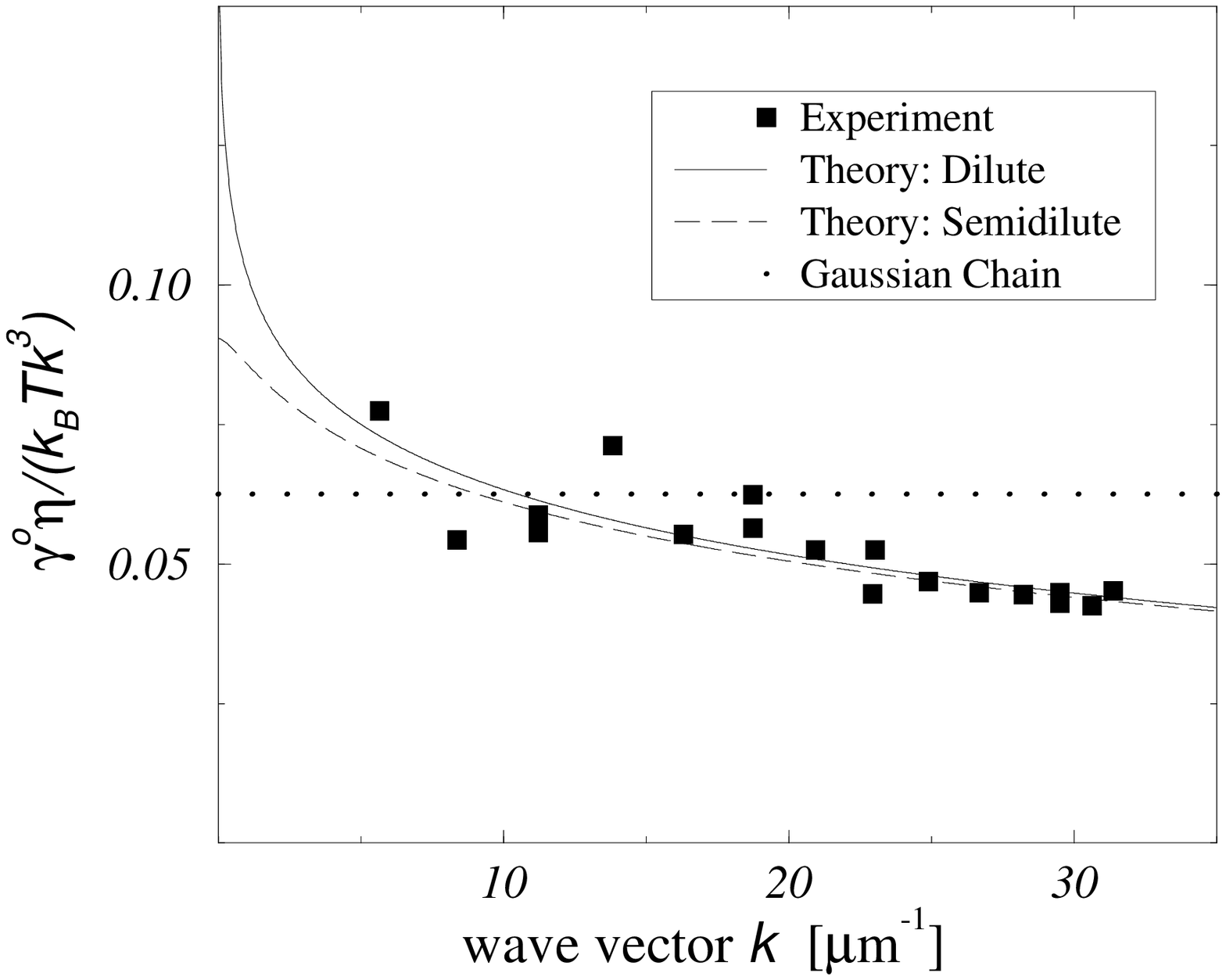}}
    \centerline{\epsfxsize=2.8truein \epsffile{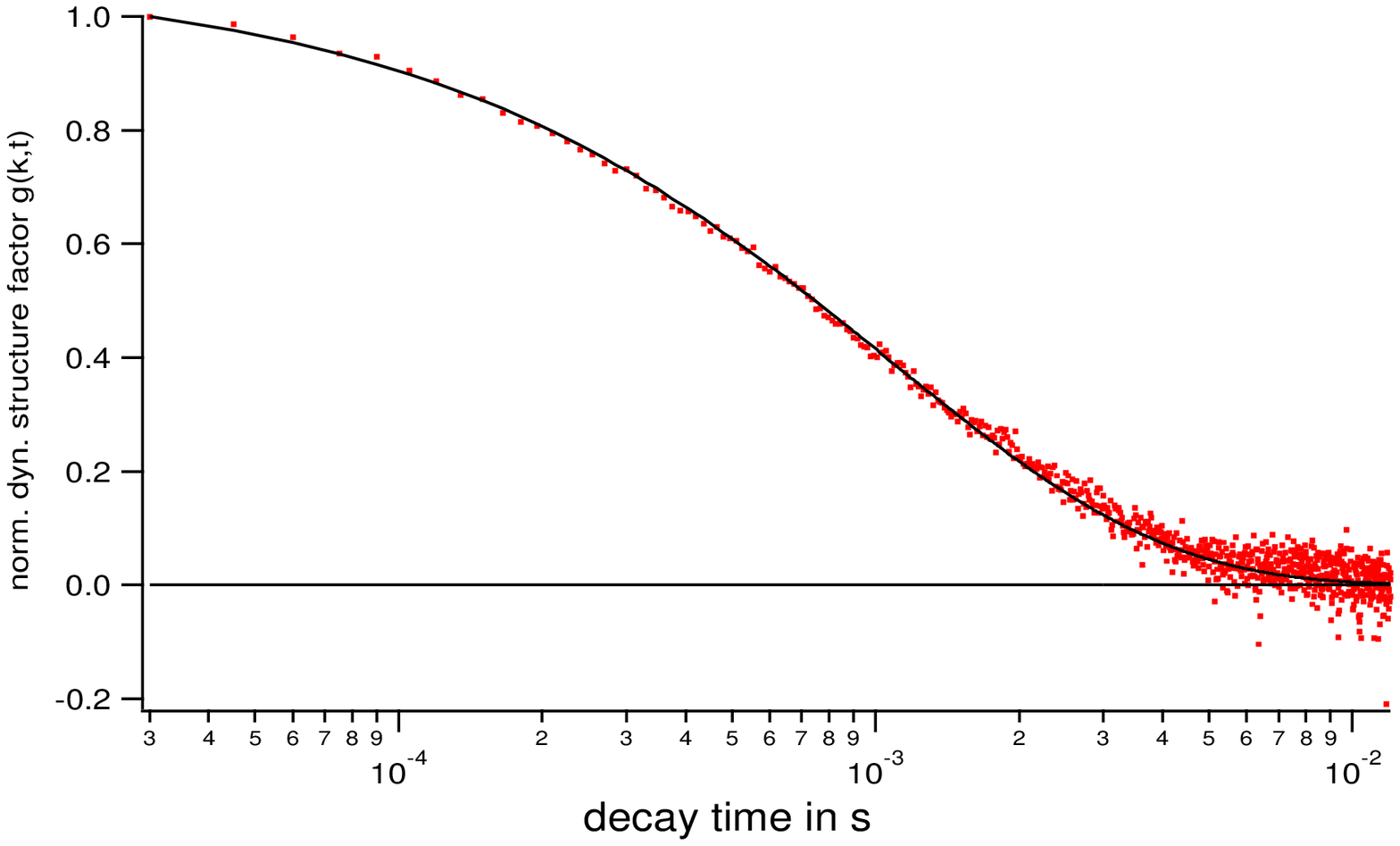}}
    \vspace{-0.5cm}
    \caption{\label{fig:initial_decay} 
      \small Results from dynamic light scattering experiments.  {\it Left:}
      Correction to the classical prediction $\gamma_k^{(0)} \sim k^3$ for the
      initial decay rate of the dynamic structure factor. The theoretical
      predictions for dilute solutions (solid line) and semidilute solutions
      (dashed line) are compared with experimental data
      \protect\cite{schmidt-etal:89}. Also included is the prediction for
      Gaussian chains from Ref.\ \protect\cite{doi-edwards:86}.  {\it Right:}
      Fit of theoretical dynamic structure factor to experimental data for $k=
      24.2 \, \mu {\rm m}^{-1}$ \protect\cite{goetter-etal:96}.}
  \end{center}
\end{figure}
The effective transversal friction coefficient per length, $\tilde\zeta_\perp$,
accounts for the hydrodynamic damping of the undulations on length scales
comparable to the scattering wavelength. For polymers with
$\ell_p\approx\lambda$ mainly the first regime, Eq.\ \ref{eq:gamma_0}, is
observed.  The microscopic lateral diameter $a$ of the polymer enters the
calculation as a microscopic cutoff and can be determined from a measurement of
the {\em initial decay} of the dynamic structure factor.  The method is easy to
apply and has been shown to provide reasonably accurate results (see
Fig.~\ref{fig:initial_decay}).  For polymers with $\ell_p\gg\lambda$ mainly the
stretched exponential decay regime, Eq.\ \ref{eq:stretched_exp} is observed.
The decay rate is determined by the persistence length of the molecule, which
is also readily extracted from experimental data. Some applications of the
methods to dynamic light scattering with actin are shown in
Fig.~\ref{fig:initial_decay}.

Our analytical results for the initial decay rate and the dynamic exponent for
semiflexible polymers suggest that known deviations of the dynamic exponent for
more flexible polymers from its classical value $z=3$ are most likely due to
the local semiflexible structure of these molecules, and that the above
analysis is therefore of some relevance also to scattering from flexible
polymers.  The reason is that the singular ($\propto 1/r$) hydrodynamic
interaction favors short distances. The possibility to determine the
microscopic lateral diameter $a$ of the polymer from the initial decay becomes
important when the effects of bundling and side-binding are investigated, which
can give rise to an effective thickening of the molecules.


\section{Many-chain properties}

As we have mentioned in the introduction, we are still far from understanding
on a microscopic basis the macroscopic viscoelastic properties of
solutions and gels of semiflexible polymers. In the following we will review
some idealized concepts we have developed to model certain aspects of this
complicated behavior and compare our predictions with some recent experiments.

\subsection{Entanglement transition}

\begin{figure}[htb]
\begin{center}
  \valign{\vfil#\vfil\cr \hbox{\includegraphics[height=1.6in]{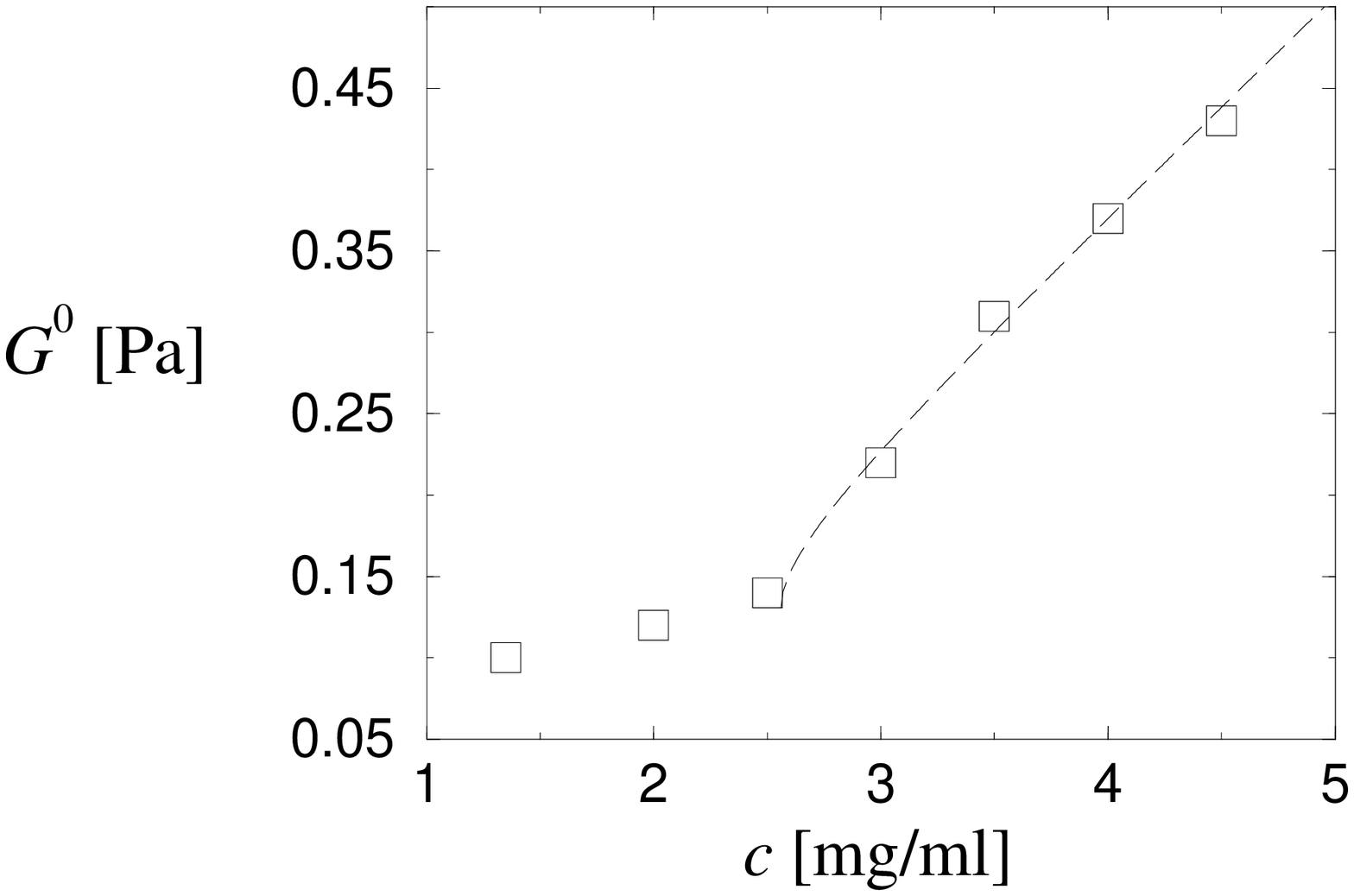}} \cr
    \hbox{\hspace*{0.5cm}
      \includegraphics[height=1.76in]{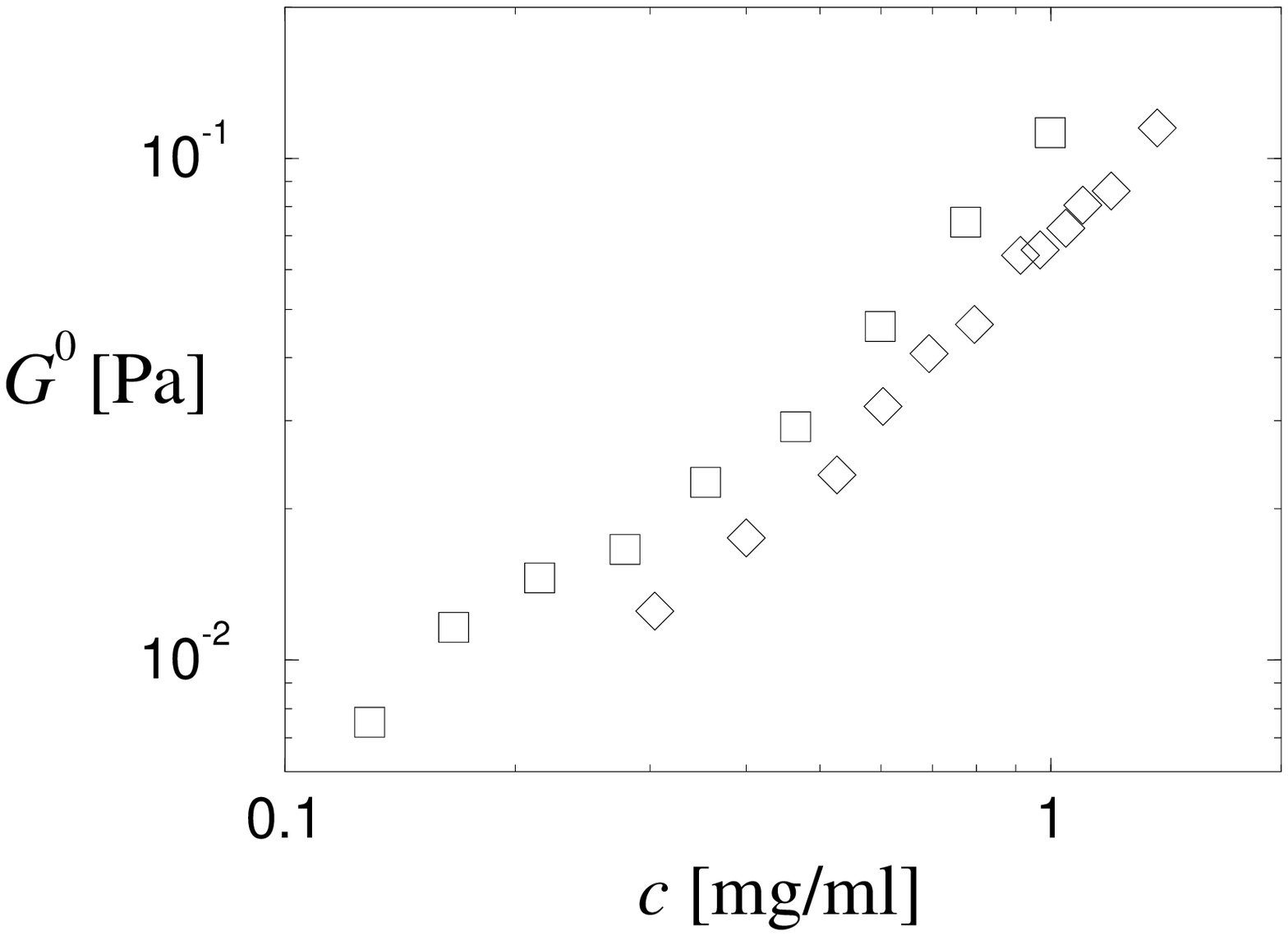}}\cr}
 \caption{\label{fig:g0_c} 
   \small {\it Left:} The entanglement transition for short rod-like actin
   filaments ($L\approx 1.5$ $\mu$m) of varying concentration.  {\it Right:}
   Concentration dependence of the plateau modulus for pure actin ($\square$)
   and actin with a small amount of gelsolin ($\diamond$)
   \protect\cite{hinner-etal:97}.}
  \end{center}
\end{figure}

Upon increasing the polymer concentration $c$ at fixed polymer length there is
a critical concentration $c^*$ above which a plateau develops in the frequency
dependence of the storage modulus and the modulus increases steeply. The same
phenomenon is observed at a critical length when the polymer length is varied
at fixed concentration. It is called the entanglement transition. An
experimental observation of the transition for short actin filaments
($L\approx1.5$ $\mu$m) as a function of concentration is shown in
Fig.~\ref{fig:g0_c} (left).  
Adapting concepts which have been developed for
flexible polymers \cite{kavassalis-noolandi:89} we have proposed a mean field
description for the entanglement transition in semiflexible polymer solutions
\cite{kroy-frey:96}.  The basic idea is that the ends of a polymer are less
efficient in confining other polymers than internal parts of the polymer. As a
consequence the critical concentration $c^*$ is related to the overlap
concentration ${\bar c}$ by some universal number $C$ which may be interpreted
as an effective coordination number and is a measure of the mean number of
neighboring polymers necessary to confine the lateral motion of the test
polymer. The prediction for an ideal gas of rods, $G^0\propto k_BT/\xi_{\rm
  eff}^2L$, is also shown in Fig.~\ref{fig:g0_c}. From the fit we have
determined the coordination number $C$, which compares very well with the
result for flexible polymer solutions.  Away from the transition the effective
mesh size reduces to the ordinary mesh size $\xi_{\rm eff}\propto \xi_m$.  With
$\xi_m\propto c^{-1/2}$ \cite{schmidt-etal:89} this implies $G^0\propto c$,
which is indeed observed for a regime of concentrations within the semidilute
phase.  However, at higher concentrations or lengths the slope of $G^0(c)$
increases, as can be seen from the bottom plot in Fig.~\ref{fig:g0_c}, which
shows the measured plateau modulus $G^0$ for pure actin and actin with a small
amount of gelsolin ($r_{\rm AG}=6000:1$ corresponding to an average actin
filament length of 16 $\mu$m).  The description of the polymers as rods, which
neglects internal modes of the individual polymers, is no longer appropriate.
NOTE: Since the time of writing this manuscript, new experimental and
theoretical investigations \cite{hinner-etal:97} have suggested another,
probably superior, interpretation of the entanglement transition, which does
not involve the concept of the coordination number.

\subsection{Plateau modulus}

\begin{figure}[tb]
\begin{center}
  \includegraphics[height=.9in]{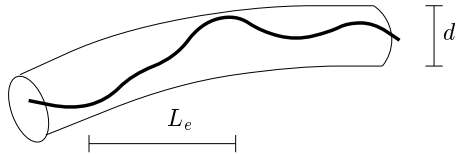}
 \caption{\label{fig:odijk} 
   \small Some of the theoretical work is based on a tube picture as depicted
   schematically here. The effect of the network surrounding an arbitrary test
   polymer is represented by a cylindrical cage of diameter $d$. Different
   scaling laws for $d$ have been proposed.}
  \end{center}
\end{figure}

To explain the observed elastic modulus on the basis of the elastic properties
of the individual chains and the architecture of the network is in general a
quite difficult task.  Some of the theoretical work is based on a tube picture
as depicted schematically in Fig.~\ref{fig:odijk}. The effect of the network
surrounding an arbitrary test polymer is represented by a cylindrical cage of
diameter $d$. The main response of the polymer to various deformations of this
tube is assumed to arise from distortions of undulations of wavelength $L_e$,
with $L_e^3\propto d^2\ell_p$ \cite{odijk:83}. It is far from obvious, whether
the macroscopic elasticity can be explained solely in terms of the
compressibility of the tubes \cite{odijk:83,isambert-maggs:96}, or whether
contributions from filament bending \cite{kroy-frey:96,satcher-dewey:96} or
buckling \cite{mackintosh-kaes-janmey:95} are also important.  The theoretical
approaches differ quite significantly in their assumptions on how forces are
transmitted through the network and which type of deformation of a single
polymer gives the dominant contribution to the network elasticity. Due to those
differences in the basic assumptions the predictions for the shear modulus of
the network also differ.
Our present data are reasonably consistent with entropy arguments assuming
$G^0\propto k_BT/\xi_m^2L_e$, which leads to $G^0\propto c^{4/3}$ for $d\propto
\xi_m$ and $G^0\propto c^{7/5}$ for $d\propto \xi_m^{5/6}/\ell_p^{1/5}$
\cite{isambert-maggs:96}, respectively.  Above 0.4 mg/ml our data also agree
with a scaling $G^0\propto c^{5/3}$, which has been derived by a scaling
argument based on Fig.~\ref{fig:odijk} for the case that the macroscopic
elastic response is mainly due to filament bending \cite{kroy-frey:96}.  It
should also be noted that these concentrations are close to the critical volume
fraction $3a/\!L$ ($a\approx 10$ nm is the lateral diameter of an actin
filament) for the nematic transition of rigid rods.

\subsection{Terminal relaxation}

At frequencies below the plateau regime the elastic response decreases and the
polymer solution starts to flow. The corresponding time scale is called the
terminal relaxation time. It can be determined from the measured plateau
modulus and the zero shear rate viscosity using the relation $\eta_0 \sim
G^0\tau_r$ or directly from $G'(\omega)$.

\begin{figure}[htb]
\begin{center}
  \valign{\vfil#\vfil\cr
    \hbox{\includegraphics[angle=-90,width=0.3\textwidth]{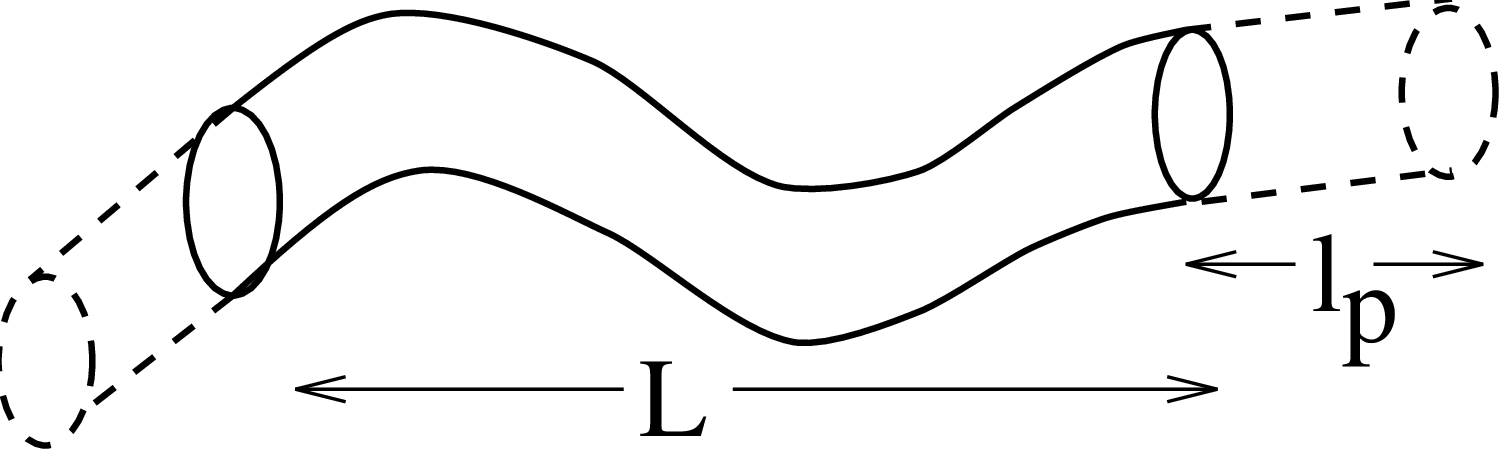}} \cr
    \hbox{\hspace{1.5cm}
      \includegraphics[width=0.5\textwidth]{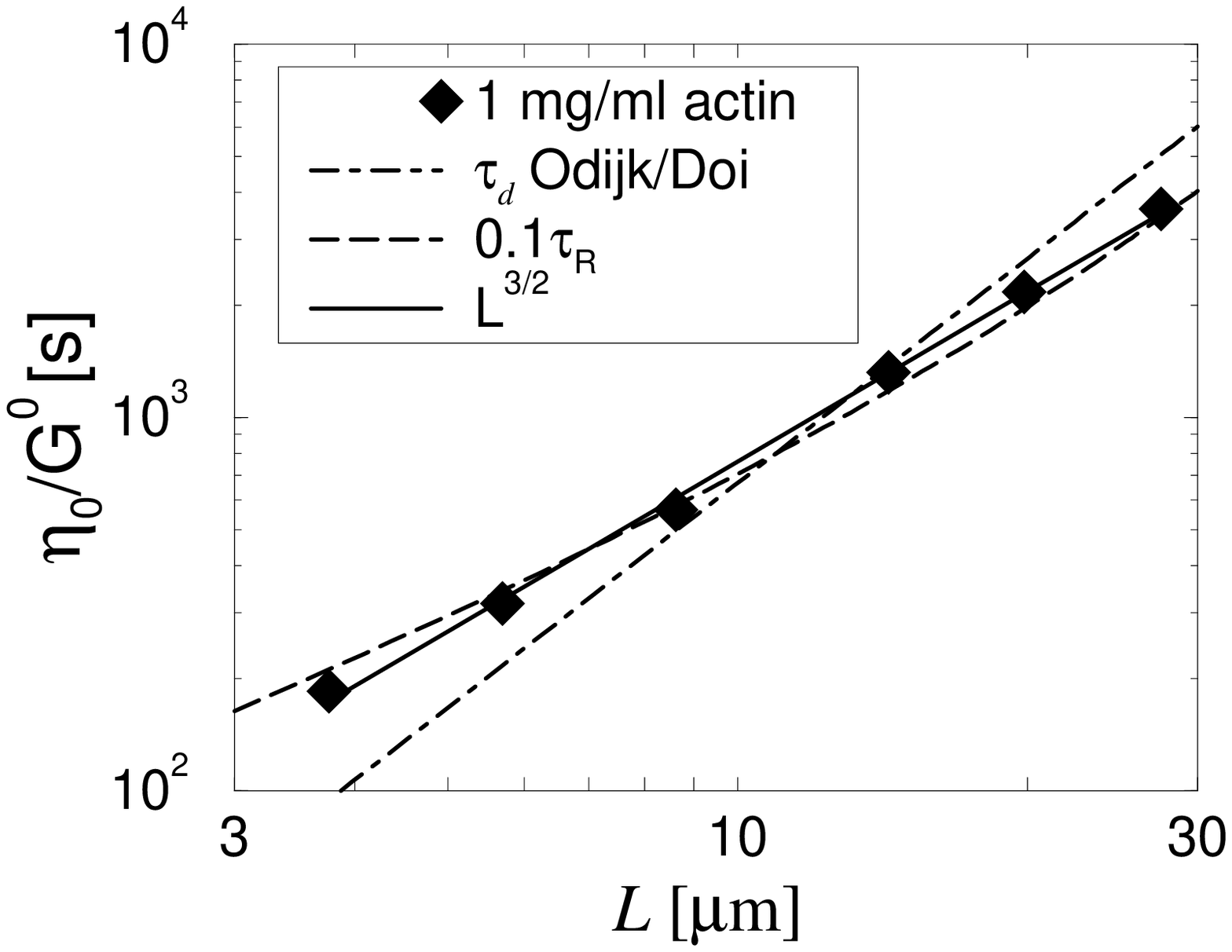}}\cr}
\end{center}
\caption{\label{fig:reptation} \small {\it Left:} 
  Illustration of the generalized reptation picture for semiflexible polymer
  solutions. {\it Right:} Terminal relaxation time above the entanglement
  transition \protect\cite{hinner-etal:97}.}
\end{figure}

Intuitively, the mechanism for the terminal relaxation is obvious from the tube
picture (see Fig.~\ref{fig:reptation}) described above: viscous relaxation only
occurs, when the polymers have time to leave their tube-like cages by
reptation, i.e., by Brownian motion along their axis.  The reptation model,
which was originally formulated for flexible polymers, was adapted to the
semiflexible case \cite{odijk:83}.  Odijk \cite{odijk:83} estimated the
disengagement time $\tau_d$ for a semiflexible chain diffusing out of its tube.
However, the data for $\tau_r$ presented in Fig.~\ref{fig:reptation} (bottom)
are not in accord with his result for $\tau_d$.  The dependence of the observed
terminal relaxation time $\tau_r$ on polymer length $L$ is substantially weaker
than predicted for $\tau_d$, even in the stiff limit where
$\tau_d=\ell_pL/4D_\| \propto L^2$ (dot-dashed line in
Fig.~\ref{fig:reptation}), $D_\| = k_BT/2\pi\eta L$ being the longitudinal
diffusion coefficient of the chain in the free draining approximation.
Instead, the solid line in Fig.~\ref{fig:reptation} corresponds to the scaling
law $\tau_r\propto L^{3/2}$.

A tentative interpretation of the data can be given in terms of a semiflexible
polymer diffusing along a strictly one dimensional path; i.e., not being
allowed to choose between infinitely many new directions at its ends. This
situation is schematically depicted in the upper part of
Fig.~\ref{fig:reptation}.  The characteristic decay time for self-correlations
of the end-to-end vector $\langle {\mathbf R}(t){\mathbf R\rangle}$ is then
given by \cite{hinner-etal:97}
\begin{equation}
  \tau_R = L^4\ell_p^2/D_\| \langle R^2\rangle^2 \approx
  (L+2\ell_p)^2/4 D_\|\;.
\end{equation}
This presents an upper bound for the terminal relaxation time within the tube
model. As seen from the dashed line in Fig.~\ref{fig:reptation}, $\tau_R$ (for
$\ell_p= 17$ $\mu$m \cite{gittes-etal:93,ott-magnasco-simon-libchaber:93}) is
in fact by a factor of ten too large compared to the data but describes fairly
well the length dependence of $\tau_r$. The restriction to one path implies a
very slow decay of conformational correlations. An unusually slow decay of
stress (the frequency dependence of $G'(\omega)$ is still less than linear in
the measured frequency range) is indeed observed, but this might also in part
be due to the broad length distribution of actin \cite{janmey-etal:86}.
Clearly, further investigations are necessary to come to a better understanding
of the terminal regime.


\subsection{Stochastic network models}
\label{stochastic_network_models}

As motivated in the introduction, we use a two dimensional toy model of
crosslinked semiflexible polymer networks to study some of the fundamental
problems in semiflexible network theory~\cite{wilhelm-frey:97b}. Here we
address the question whether longitudinal or transversal deformation dominate
the network response and find an interesting crossover between the two extreme
cases. Since such questions are already nontrivial for networks at $T=0$ (i.e.
no fluctuations), we consider this most simple case first: The linear elastic
response of a network of classical compressible and bendable rods. A more
sophisticated treatment will include contour fluctuations of the filaments (but
no steric interactions) by replacing the rods with elements having the
appropriate linear or nonlinear force extension relations described above.

Rods of unit length are randomly placed on a square piece of the plane.
Wherever two sticks intersect, they are connected by a crosslink. Periodic
boundary conditions are applied to the left and right sides of the square. A
shearing horizontal displacement is enforced on all rods intersecting the upper
or lower boundary of the square at the intersection points. The (linearized)
elastic equations of the system are solved numerically for displacements and
forces at all crosslink points. The shear modulus is obtained from the force
needed to impose the displacements of the upper and lower boundary.  Crosslinks
are not allowed to stretch but do not fix the angle between the intersecting
rods. The results reported here change quantitatively but not qualitatively
when crosslinks do fix the intersection angles.

The parameters of this system are the density $\rho$ of rods per unit area,
the compressional and the bending stiffness of the rods.  We choose to
express the latter by the linear response force constants $k_{\text{comp}}$
and $k_{\text{bend}}$ for a rod of unit length. Geometrically, the relative
size of the two force constants is controlled by the aspect ratio $\alpha =
a/L$ of the rods where $a$ is the rod radius.
\begin{figure}[bt]
  \centerline{\includegraphics[height=0.4\textwidth]{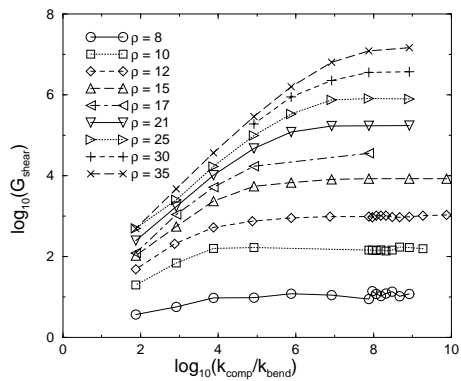}}
    \caption{\label{fig:stochastic_models}
      \small Shear modulus of a crosslinked two dimensional random network of
      rods of unit length for different densities $\rho$ and different aspect
      ratios $\alpha$ plotted over the relative size
      $k_{\text{comp}}/k_{\text{bend}}$ of compressional and bending force
      constants. Units were chosen in such a way that the bending stiffness
      $\kappa$ and hence $k_{\text{bend}}$ were constant ($\kappa = 1$).}
\end{figure}

While the system can show no elastic response (in the limit of infinite size)
for densities below the percolation threshold $\rho_c^g \approx 5.72$
\cite{pike-seager:74a}, the shear modulus $G$ grows very rapidly with
increasing $\rho$ for $\rho \geq \rho_c$. For sufficiently slender rods $G$
displays unusual power law behavior in $\rho$ even for denisties far above the
percolation threshold.

To address the question whether the elasticity of a random stiff polymer
network is dominated by transverse or by longitudinal deformations of the
filaments, one can study the dependence of the shear modulus on the two force
constants.  We keep $k_{\text{bend}}$ fixed and increase $k_{\text{comp}}$ from
values corresponding to $\alpha = 0.15$ (short, thick rod) to values
corresponding to $\alpha = 1 \times 10^{-5}$ (long slender rod) for different
system densities (see Fig.~\ref{fig:stochastic_models}). We observe that beyond
a certain point the shear modulus ceases to depend on $k_{\text{comp}}$,
indicating that the elasticity is dominated by bending modes for thin rods.
Since the relevant scale for the system elasticity at higher densities is set
not by the rod length but by the mesh size, the point of onset for this
behavior shifts upwards with density. The dominance of bending modes in this
region is confirmed by the observation that almost all of the energy stored in
the deformed network is accounted for by the transverse deformation of the
rods.  Even in the region were the shear modulus does depend on
$k_{\text{comp}}$, a substantial part of the elastic energy is stored in
bending deformations. \\
{\bf Acknowledgment:} Our work has been supported by the Deutsche
Forschungsgemeinschaft (DFG) under Contract No.\ SFB 266 and No.\ Fr 850/2.


\end{document}